\documentclass{iau} 
\usepackage{graphicx}
\usepackage{caption}
\usepackage{subcaption}

\newcommand{\aap}{{\it A\&A}}

\title[Kepler-71 Activity]{Analysis Of Kepler-71 Activity Through Planetary Transit}

\author[E.~A.~Gusm\~ao \& C.~L.~Selhorst \& A.~S.~Oliveira]{Eber A. Gusm\~ao$^1$, Caius~L.~Selhorst$^{1,2}$ \and Alexandre S. Oliveira$^1$}

\affiliation{$^1$IP\&D - Universidade do Vale do Para\'iba - UNIVAP \\ S\~ao Jos\'e dos Campos, SP, Brazil \\ email: {eber.gusmao@hotmail.com\\
alexandre@univap.br} \\[\affilskip]
$^2$NAT - N\'ucleo de Astrof\'isica Te\'orica – Universidade Cruzeiro do Sul \\ S\~ao Paulo, SP, Brazil \\ email: {caiuslucius@gmail.com}}

\pubyear{2016}
\volume{328}  
\jname{Living Around Active Stars}
\editors{D. Nandi, A. Valio \& P. Petit, eds.}
\begin{document}

\maketitle

\begin{abstract}
An exoplanet transiting in front of the disk of its parent star may hide a dark starspot causing a detectable change in the light curve, that allows  to infer physical characteristics of the spot such as size and intensity. We have analysed the Kepler Space Telescope observations of the star Kepler-71 in order to search for variabilities in 28 transit light curves. Kepler-71 is a star with 0.923~$M_\odot$ and 0.816~$R_\odot$ orbited by the hot Jupiter planet Kepler-71b with radius of 1.0452~$R_J$. The physical parameters of the starspots are determined by fitting the data with a model that simulates planetary transits and enables the inclusion of spots on the stellar surface with different sizes, intensities, and positions. The results show that Kepler-71 is a very active star, with several spot detections, with a mean value of 6 spots per transit with size 0.6 $R_P$ and 0.5 $I_C$, as a function of stellar intensity at disk center (maximum value).

\keywords{Stars: activity - Starspots - Planetary Systems}
\end{abstract}

\firstsection 
\section{Introduction}

More than 2,000 years ago, sunspots had already been reported by the Chinese, but their scientific study began with the advent of the telescope. The spots on the surface of the Sun were first observed with the aid of a telescope by Galileo four centuries ago. The sunspots are colder regions in the photosphere with a strong concentration of magnetic field lines. Furthermore, sunspots are important signatures of the cyclic nature of the star's magnetic field and wealth of information about solar activity. It is considered that other stars also are subject to the same magnetic activity. Nevertheless, nowadays it is not possible to observe or even monitor similar spots on the surface of other stars due to their size and distance.


When a planet moves in front of its parent star and is seen by an observer, the event is called transit. Through the continuous monitoring of these transits, it is possible to study the exoplanet eclipsing its parent star and if a dark stellar spot were occulted, a detectable variation in the light curve (positive variation) can be observed as shown in Fig. 1. From modelling of these transits, it might be possible to infer the physical properties of the spots, such as size, intensity, position, and temperature (e.g., \cite{Adriana2003}, \cite{Adriana2010}).

In this work, we have analysed 28 transit light curves  of the star Kepler-71 in order to search for the starspot physical parameters and their variabilities.

\begin{figure}[!h]
   \centering
   \includegraphics[width=13.0cm]{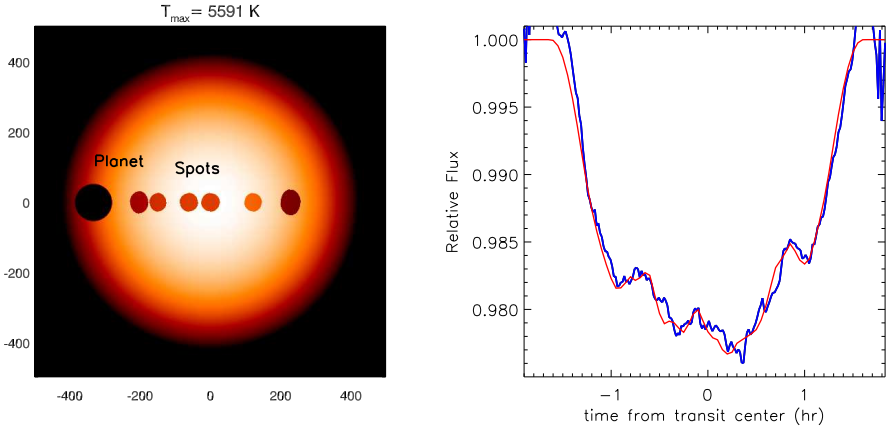} 
   \caption{Left panel: Simulated image of the star Kepler-71 with quadratic limb darkening, 6 spots, and its planet, a hot Jupiter planet Kepler-71b with radius of 1.0452~$R_J$, assumed as a dark disk. Right panel: Observed light curve (blue) and the simulated transit light curve (red).}
   \label{Fig_01}
\end{figure}

\section{Data Analyses and Results}

\begin{figure}[!h]
   \centering
   \includegraphics[width=13.0cm]{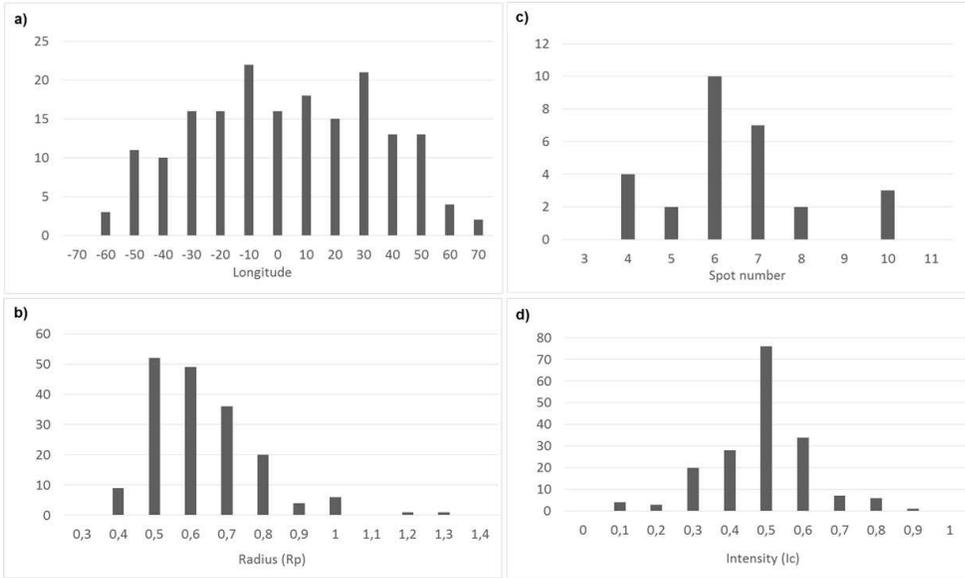} 
   \caption{Histograms of the spot parameters obtained from the fits to the light curve transits: a) longitude in the stellar disk, b) radius in units of $R_P$, c) spot number per transit and d) intensity in units of $I_C$.}
   \label{Fig_02}
\end{figure}

A planet around the star Kepler-71 was detected during one of the long run observations of a field toward the Galactic center performed by the Kepler satellite. A total of 28 transits were detected in the light curve with a high temporal resolution of $58~s$, for a total of 144~days. Kepler-71 is a star with 0.923~$M_\odot$ and 0.816~$R_\odot$ orbited by the hot Jupiter (Kepler-71b) with a mean distance of $0.047~AU$ and a  radius of 1.0452~$R_J$.

The limb darkening of the star as well as the physical characteristics of starspots are obtained by fitting the model described in \cite[Silva 2003]{Adriana2003}. 
The stelar limb darking parameters was simulated with $w_1$=0.9 and $w_2$=0.9. The round spots are modeled by three parameters: (i) intensity, as a function of stellar intensity at disk center, $I_C$ (maximum value); (ii) size, or radius, measured in units of planet radius, $R_P$; and (iii)   longitude. 

The figure 2 represents the synthesized star with spots of varying intensity (with respect to $I_C$) and radius (in units of $R_P$) and longitude. A histogram of the spot longitudes is shown in the Fig. 2 (a). These are topocentric longitudes, that is, they are not the ones located on the rotating frame of the star, but rather are measured with respect to an external reference frame. The number of spots for each transit is shown in Fig. 2 (c) and varied in the range from 4 to 10 with a mean value of 6. For spot radius, the distribution of spot radius obtained from the fits to all transit data is shown in Fig. 2 (b). The results show that the radius of the modelled spots varies from 0.4 to $1.3 ~R_P$ with a mean value of 0.6 $R_P$. Spots with lower intensity values, or higher contrast spots, are spots cooler than those with intensity values close to $I_C$. The spot intensities obtained from the model are shown in Fig. 2 (d). The figure shows that the spot intensities range from 0.1 to $0.9~I_C$  with a mean value of 0.5 $I_C$.

\section{Conclusion}

The star was evaluated using a model having up to 10 spots at a given time on its surface at certain location (latitude and longitude) during 144 days of observation by Kepler Telescope. During this period a total of 28 transits were detected. Kepler-71 is a very active star, and many intensity variations were identified in each transit (see Fig. 2), implying that there are many spots present on the surface of the star at any given time. The results show spot detections, with a mean value of 6 spots per transit with size 0.6 $R_P$ and 0.5 $I_C$, as a function of stellar intensity at disk center (maximum value). The spots on Kepler-71 has diameters in a rough order of magnitude of 44.000 km. The mean surface of star area covered by spots within the transit latitudes is in the range of 40\%. It was observed that most of spots are smaller than the planet Kepler-71b. The values obtained here can be compared with the star CoRoT-2 (\cite[Silva-Valio et al 2010]{Adriana2010}). Both star presents high activity considering the number of spots as well as spot size.

\begin{acknowledgements}
E.A.G. acknowledges a CAPES scholarship. C.L.S. acknowledge financial support from the S\~ao Paulo Research Foundation (FAPESP), grant number 2014/10489-0. 
\end{acknowledgements}


\end{document}